\documentclass[12pt]{amsart}
\usepackage[margin=1.0in]{geometry}
\usepackage{mathptmx,color,amssymb,graphicx}
\usepackage[foot]{amsaddr}
\begin{document}
\title{Spatial analysis of U.S. Supreme Court 5-to-4 decisions}
\author{Noah Giansiracusa$^*$} 
\address{$^*$Assistant Professor of Mathematics, Swarthmore College} 
\author{Cameron Ricciardi$^{**}$} 
\address{$^{**}$Undergraduate Mathematics and Economics Major, Swarthmore College}
\email{ngiansi1@swarthmore.edu, criccia2@swarthmore.edu}

\maketitle

\begin{abstract}
While the U.S. Supreme Court is commonly viewed as comprising a liberal bloc and a conservative bloc, with a possible ``swing vote'' or ``median justice'' between them, surprisingly many case decisions are not explained by this simple model.  We introduce a pair of spatial methods for conceptualizing many 5-to-4 voting alignments that have occurred on the Court and which defy the usual liberal/conservative dichotomy.  These methods, utilizing higher order Voronoi diagrams and halving lines, are based on the metric geometry of the two-dimensional ideal space locations obtained from applying multidimensional scaling to voting data.  We also introduce a two-dimensional metric method for determining the crucial ``fifth'' vote in each 5-to-4 ruling and for determining the median justice in any collection of terms within a natural court.
\end{abstract} 


\section{Introduction}

A common perspective of the U.S.\ Supreme Court, dominant in both scholarly work from the last several decades as well as popular and journalistic accounts, is that each of the nine justices serving during a specified term is placed along a one-dimensional spectrum measuring in some manner the liberal-to-conservative outlook of the justice.  A typical situation is that the four most liberal justices are viewed as a voting bloc, as are the four most conservative, and the unique justice lying between these on the one-dimensional scale is considered a powerful ``median justice,''  capable of joining either bloc and thus in essence deciding the outcome of every politically contentious case with his or her ``swing vote.''\footnote{This idea is so pervasive and taken to such extents that experts have advocated the following strategy: to win cases being argued before the Supreme Court, one should identify the median justice and form the argument that maximizes the chances of obtaining that justice's swing vote (see \cite[p.14]{Controversial5-4} and \cite[p.217]{SandraDayMostPowerful}).}  For instance, the 2009--2015 natural court is widely considered to have a conservative bloc consisting of Alito, Roberts, Scalia, and Thomas, and a liberal bloc consisting of Breyer, Ginsburg, Kagan, and Sotomayor, with the swing vote provided by Kennedy \cite{KennedyQuote,SuperMedians}.

While many case outcomes do neatly align with this one-dimensional picture, many do not and so appear surprising, unusual, perplexing, disordered, etc.  A highly salient example occurring during the natural court mentioned above is the 2012 case \emph{National Federation of Independent Businesses v.\ Sebelius}, where Kennedy swung to the conservatives but Roberts shocked many Supreme Court observers by defecting and providing the liberals with a crucial win in upholding the individual mandate of the Affordable Care Act on the grounds of the Constitution's Taxing and Spending Clause.  The one-dimensional interpretation of this case is simply that Roberts voted more liberally than expected and thus upset the traditional balance of the court.  In the present paper we introduce several mathematical methods for analyzing Supreme Court voting patterns, by way of planar geometry, in order to provide deeper insight into cases such as this one and other 5-to-4 decisions.


\subsection*{Acknowledgements}

We thank Paul Edelman for conversations on Supreme Court data and the literature analyzing it, and Michael Burr for conversations on computational geometry.


\section{Background and prior work}

The field of quantitative/empirical legal studies has been growing at a remarkable rate as data sets and data analysis methods continue to improve \cite{DefenseEmpirical,EmpiricalIntro,QuantPred,SegalSpaeth,CodingComplexity}.  Within this vast body of literature are certain papers that apply deeply mathematical tools to study the behavior of the U.S. Supreme Court.  For instance, in recent years we have witnessed the application of: information theory to determine the number of dimensions the justices vote in \cite{PNAS-2003-Sirovich}; statistical physics to measure the influence justices exert on each other \cite{StatMechSC}; game theory to compute the voting power of each justice \cite{EdelmanChen}; machine learning to predict the outcome of cases \cite{SC-ML,SupCourtML}; Bayesian methods to estimate the political ideology of the justices \cite{MartinQuinn}.  While statistics drives these studies and plays an implicit role in the present paper as well, one of our goals is to introduce more \emph{geometric} methods into this body of research.   Since our focus is on 5-to-4 to decisions and swing votes and how these interact with the dimensionality of the Supreme Court, we begin with a discussion of previous investigations into these topics.


\subsection{5-to-4 votes and oddball coalitions}

Given that a fully staffed Supreme Court comprises nine sitting justices, the potential for any group of five of them to form a coalition is extremely important and powerful.  The following amusing piece of legal folklore concerning Justice William Brennan, who served from 1956 until 1990, aptly captures this phenomenon:
\begin{quote}
At some point early in their clerkships, Brennan asked his clerks to name the most important rule in constitutional law. Typically they fumbled, offering \emph{Marbury v.\ Madison} or \emph{Brown v.\ Board of Education} as their answers. Brennan would reject each answer, in the end providing his own by holding up his hand with the fingers wide apart. This, he would say, is the most important rule in constitutional law.  Some clerks understood Brennan to mean that it takes five votes to do anything, others that with five votes you could do anything. \cite[p.763]{RuleofFive-anecdote}
\end{quote}
Despite the recognition of their significance,\footnote{And prevalence: Enns and Wohlarth wrote in 2013 that, ``Since 1946, 17\% of all Supreme Court cases---and 37\% of `landmark decisions'---have been decided by one vote'' \cite[p.1101]{SwingJustice}.} much remains to be understood about the formation of coalitions of five---or, to put it another way, about the distribution of 5-to-4 voting patterns among the nine justices.  For instance, as Edelman astutely points out, the above-cited 2003 paper that uses information theory to study Supreme Court voting patterns \cite{PNAS-2003-Sirovich} also uses singular value decomposition to approximate voting behavior yet ``it is on the 5--4 decisions other than the c[onservative]-l[iberal] split that this approximation performs at its worst, which is not that surprising given that these are the decisions for which no one has a good theoretical explanation'' \cite[p. 569]{DimSupCourtEdelman}.   

In 1993 Riggs published an in-depth study of 5-to-4 decisions, focusing primarily on the frequency of such cases and how this interacts with the ideology of the Court \cite{5-4Decisions}.  Riggs notes at the time of his writing that ``no systematic study of this voting pattern over an extended time period has previously been published'' \cite[p.669]{5-4Decisions}.  He found a significant increase in the frequency of 5-to-4 decisions throughout the 20th century.  In 2014, Friedland published a study of 5-to-4 decisions focusing on the context of public response and the politicization of the Court \cite{Controversial5-4}.  He notes in passing that ``many cases can not be seen as dividing along liberal/conservative lines'' \cite[p. 43, note 169]{Controversial5-4}.  So we are left in awkward situation: as the Brennan folklore illustrates, majorities of size five are fundamental to Supreme Court jurisprudence, yet as Friedland points out many such majorities are not the ones we expect and as Edelman emphasizes, these are precisely the cases for which current theories are most deficient.  

For many years the \emph{Harvard Law Review} has published data on Supreme Court rulings, including a tabulation of 5-to-4 decisions listing the number of decisions made by each coalition of five justices.  Cases resulting in 5-to-4 decisions by particularly striking coalitions have been studied by many scholars, for example \emph{National Federation of Independent Business v.\ Sebelius} \cite[p.43]{Controversial5-4}, \emph{Rogers v.\ Tennessee} \cite[p.819]{EdelmanMeasuringDeviations}, \emph{Philip Morris v.\ Williams} \cite[p.820]{EdelmanMeasuringDeviations}, \emph{Irizarry v.\ United States} \cite[p.1092]{SwingJustice}, \emph{Kyllo v.\ United States} \cite[p.35]{CoalitionFluidity}, to name just a few.  Friedland reviewed 5-to-4 decisions, emphasizing both individual cases and overall trends \cite[p.17]{Controversial5-4}, and in 2016 Fischman and Jacobi discussed a particular ``unusual'' five-justice coalition that occurred several times\footnote{Since the term ``unusual'' suggests rarity, which evidently is not accurate, we prefer the term ``oddball'' coalition used by Edelman and Chen \cite[p.68, note 24]{EdelmanChen}.} \cite[p.1673]{2ndDimension}.  

In 2008, Edelman, Klein, and Lindquist defined a measure of the ``disorder'' of a judicial coalition, indicating how far it is from the expected liberal/conservative divisions \cite{EdelmanMeasuringDeviations}.  We shall discuss this further below in \S\ref{sec:disorder}---for now,  we quote one of their intriguing findings concerning the most disordered cases among the terms they analyze:
\begin{quote}
Many (but not all) of these disordered decisions involve legal questions that are often very technical and, in that sense, less clearly ``ideological.'' First, most involve statutory rather than constitutional interpretation, including, for example, the Internal Revenue Code, the Antitrust Act, the Labor Management Relations Act, the Hobbs Act, the Bankruptcy Code, the Freedom of Information Act, the postal and mail fraud statutes, and the Copyright Act. Second, they often involve procedural rather than substantive matters, including service of process, jurisdiction, counterclaims, summary judgment, and harmless error on appeal. \cite[p.833]{EdelmanMeasuringDeviations}
\end{quote}
In sum then, it seems that mathematical and statistical methods have been used to highlight individual cases decided by oddball coalitions and to convey overall trends in 5-to-4 decisions, but for conceptualizing such cases and voting patterns scholars mostly rely on traditional direct legal analysis.


\subsection{Ideal points, multidimensional scaling, and the second dimension}

The notion of ``ideal points'' has been rather influential in political science, including Supreme Court scholarship.  In brief, in the context of the Supreme Court this refers to coordinatizing justices and/or cases in some kind of ideology or policy space, invariably taken to be Euclidean space $\mathbb{R}^d$ for some $d > 0$.  The most ubiquitous instance of this is assigning a real number (so $d=1$) to each justice representing their location on a liberal-to-conservative spectrum.  In practice, there are various ways of accomplishing this.  The ``Segal-Cover scores'' are based on data external to the Court, such as newspaper editorials \cite{SegalCover}; the Bayesian ``Martin-Quinn scores,'' which use Markov chain Monte Carlo methods to dynamically compute ideal points from longitudinal data, have been a very successful and popular improvement \cite{MartinQuinn}; Peress offers another statistical approach using additional structure oriented toward the fact that the number of ideal points being estimated for the Supreme Court is relatively small \cite{Peress-IdealPoint}.  

Another approach to computing ideal points is multidimensional scaling (MDS).  Given a symmetric $n\times n$ matrix of nonnegative numbers, thought of as distances (or approximations thereof) between $n$ objects, and a positive integer $d$, MDS produces $n$ points in Euclidean space $\mathbb{R}^d$ such that their pairwise distances are close to the entries of the input matrix \cite{MDS2,MDS1}.  To use MDS to compute Supreme Court ideal points, scholars typically\footnote{Schubert largely initiated this approach in his seminal 1965 treatise \cite{Schubert}, but Brazill and Grofman in 2002 popularized, modernized, and refined it, while also comparing it (favorably) to related methods such as factor analysis \cite{MDS-vs-factor}.} take the $ij$ entry of the $9\times 9$ input matrix to be one minus the fraction of cases during a given natural court that Justice $i$ and Justice $j$ voted ``together''---meaning either both with the majority or both against the majority\footnote{This means the disposition of the decision is what matters rather than the opinion, so for instance if two justices ultimately reach the same conclusion but for different reasons and consequently write separate opinions, they are nonetheless counted as voting together.}---and a widely used way of accessing this information, which we use in the present paper as well, is the binary ``majority'' variable in the Spaeth Supreme Court Database \cite{SCDB}.  While the exact coordinates of ideal points vary across the different estimation methods, in one dimension ($d=1$) the order of the nine justices for each natural court is remarkably consistent across the different methods and is generally compatible with expert opinion of the political leanings of the justices \cite[pp.1693--4]{2ndDimension}---so it is safe to view the first dimension in all these methods as quantifying the liberal-to-conservative spectrum.\footnote{A word of caution: MDS is only well-defined up to coordinate reflections, so liberal-to-conservative may show up as either left-to-right or right-to-left, but it is usually straightforward to determine which is which and hence to fix the direction.}

But what of ideal points in higher ($d > 1$) dimensions?\footnote{For a list of papers questioning the common assumption of a one-dimensional court, see \cite[pp.1675--6, note 21]{2ndDimension}.}  For Segal-Cover scores, one would need to directly define these additional dimensions and somehow estimate them; however, for ideal point estimation methods such as Martin-Quinn and MDS that rely on voting data, we can compute coordinates in higher dimensions without having an  \emph{a priori} interpretation of the higher dimensions.  The standard MDS algorithm is inductive in the sense that it first finds the optimal coordinates in the first dimension, then it fixes these and moves the points in the second dimension to improve the distance approximations, etc., so that if $d' < d$ then the coordinates for the $d'$-dimensional MDS are the projection onto the first $d'$ components of the coordinates of the $d$-dimensional MDS.  This means that it is a well-defined, and indeed fascinating, question to ask what each dimension beyond the first of the MDS-based ideal points of Supreme Court justices represents in a jurisprudential sense.  

The pioneering work of Schubert in the 60s and 70s on applying MDS to voting records was multi-dimensional \cite{Schubert, Schubert2}, though limits to computational capabilities at the time were significant.  In 2002, Grofman and Brazill found that, ``when properly used, MDS does not normally create artifactual additional dimensions (e.g., an extremism dimension) the way that factor analysis inevitably does when applied to attitudinal data or to data on voter choices or preferences that has been generated by spatial proximity in unfolding terms'' \cite[p.56]{median-justice-MDS}.  Hook in 2007 depicted the $d=2$ MDS coordinates for nine justices but did not attempt to interpret the second dimension \cite[p.254]{StatisticalVisual}.  Peress in 2009 admits that interpreting the second dimension of his scores is a ``difficult task,'' but he finds ``some evidence to suggest that the second dimension captures `judicial activism''' \cite[p.286]{Peress-IdealPoint}.

Arguably the most in-depth analysis of the second MDS dimension of the Supreme Court is the aptly named 2016 paper by Fischman and Jacobi, which mostly focuses on the Roberts Court \cite{2ndDimension}.  They point out that since MDS is based on voting agreement rates it obviates the need to determine the ideological direction of rulings, a notoriously subtle and subjective endeavor that has continually plagued quantitative Supreme Court analysis.  They also note that the planar layout of the justices determined by MDS for the 2005--2008 terms is quite similar to that of the 2010--2012 terms (among the justices serving on both), suggesting a certain robustness to MDS-based estimation of ideal points.  

Fischman and Jacobi use the Edelman-Klein-Lindquist disorder formula to compute two separate disorder scores for each case, one for the projection of the $d=2$ MDS ideal points onto the first component and one for the projection onto the second component.  They then study the cases that have the most imbalanced scores in the sense that they are highly disordered in the first dimension but highly ordered in the second dimension---concluding that the second dimension should be interpreted primarily as measuring legal methodology \cite[p.1675]{2ndDimension}, specifically ``pragmatism versus legalism'' \cite[p.1709]{2ndDimension}.  

While this work of Fischman and Jacobi is a significant step forward in embracing and understanding the second dimension, it curiously seems to mostly avoid studying the rich interactions between the two dimensions,\footnote{An important exception  however, and possibly their motivation for treating the two dimensions separately, is their observation that ``many of the disagreements within each bloc are orthogonal to the disagreements between the blocs'' \cite[p.1679]{2ndDimension}.} primarily adopting the perspective that cases are typically determined by the linear ordering of the justices in the first dimension but for cases which are not, the next step is to consider the linear ordering of the justices in the second dimension.  The authors assert that ``the second dimension also explains many of the coalitions that cannot be explained by the first dimension,'' \cite[p.1679]{2ndDimension} so we see significant progress in addressing the concern Edelman raised in 2004 that oddball coalitions are lacking in theoretical explication \cite[p.569]{DimSupCourtEdelman}.  However, this treatment of the two dimensions as independent manifestly leaves open the door for a deeper analysis of the full planar geometry of $d=2$ MDS-based ideal points, moving beyond the two one-dimensional projections.\footnote{For those who have studied algebraic geometry, this is somewhat analogous to the difference between the Zariski topology on $\mathbb{A}^2$ and the product topology coming from the Zariski topology on $\mathbb{A}^1$.}  Very simply put, one could look at lines of arbitrary slope and also circles in the two-dimensional MDS space, rather than just vertical and horizontal lines; this is largely the starting point for the present paper.


\subsection{Swing votes and median justices}\label{sec:backgroundswing}

The intuitive idea of the ``median justice'' is that since the Supreme Court has nine justices, the one who sits roughly in the middle, in an appropriate sense, has tremendous power because he or she can cast the decisive fifth vote in any case that otherwise splits evenly among the liberal and conservative blocs.  A median justice who tends to vacillate between these two political blocs is a ``swing vote,'' a designation associated with great significance (and a behavior associated with great consternation).  As important as both of these concepts are, in practice they are rather subtle---nuances in the definitions and in quantitative methods for measuring them can lead to varied outcomes and interpretations---and so they continue to receive considerable attention in the empirical legal studies literature.

In 1990, Blasecki published a paper using statistical methods to determine whether Justice Powell, who was frequently referred to by the press and by other legal scholars as a swing vote, should in fact be considered a swing vote \cite{PowellSwing}.  Blasecki helpfully included a thorough discussion of prior definitions of the term ``swing vote'' appearing in the scholarly literature \cite[pp.532--4]{PowellSwing}; she then proceeded to introduce a new quantitative definition, based on Guttmann scaling and bloc analysis, and determined that Powell too frequently sided with the conservative justices to be considered a swing vote overall.  Powell was alleged to be most ideologically independent on issues of civil liberties, so Blasecki focuses on civil liberties cases and concludes that the 1972--74 Courts show no pivotal position, whereas in 1975--77 Justice White has broken off and becomes a pivotal swing vote---and Justice Powell came closest to being a true swing vote in 1986 but even then his votes were too frequently aligned with the conservative bloc \cite[pp.539--40]{PowellSwing}.

Grofman and Brazill in 2002 computed the $d=1$ MDS ideal point coordinates for the justices in each natural court and determined the median justice with regard to the left-to-right ordering of these points \cite{median-justice-MDS}.  Fischman and Jacobi, who had used $d=2$ MDS, note that ``In two dimensions, different patterns of coalitions emerge: in the second dimension, it is the Chief Justice [Roberts] and Justice Sotomayor, not Justice Kennedy, who sit at the median of the Court'' \cite[p.1671]{2ndDimension}.  By this they mean that when projecting the MDS coordinates onto their second components, one obtains a linear ordering with a different median than when projecting onto the first components.  In particular, they find Roberts at the median of the Court's second dimension in the case \emph{National Federation of Independent Business v.\ Sebelius} \cite[p.1685]{2ndDimension}, which helps explain the oddball majority leading to that 5-to-4 decision, though it still leaves open the question of why the second dimension was so dominant in that case.

Martin, Quinn, and Epstein in 2005 studied median justices by using Martin-Quinn scores instead of MDS and further progressed our understanding of the concept from both a theoretical and an empirical standpoint \cite{MedianJustice-Martin-Quinn-Epstein}.  Epstein and Jacobi introduced in 2008 the notion of a ``super median,'' a justice whose vote strongly influences (or at least correlates with) the majority vote\footnote{This is related to, but distinct from, the ``voting power'' of a justice as measured by Edelman and Chen \cite{EdelmanChen}.} \cite{SuperMedians}.  One of their key insights is that the distances between the ideal points of the justices, not just the relative orderings, are crucial.  For instance, they found that Kennedy was a super median in 2006, and they posit that this is because he not only was the fifth justice when ordered by ideology, but there is a large ``gap'' between him and his ideological neighbors, Souter to his left and Alito to his right \cite[p.41]{SuperMedians}.  They identify the following super medians: Clark in 1959, Goldberg in 1962, White in 1971 (quite strongly) and 1982, 1983, 1987 and 1988, Powell in 1986, O'Connor in 1999, but the most powerful in their paper is Kennedy in 2006 \cite[p.67]{SuperMedians}.

The popular view of Sandra Day O'Connor as a powerful swing vote was explored in 2004 by Lowenthal and Palmer.  They found that she had the highest percentage among the nine justices of voting with the majority in 5-to-4 decisions from 1994 to 2002 (76.9\%), followed by Kennedy (74.4\%) \cite[p.230]{SandraDayMostPowerful}.  She wrote roughly the average number of majority opinions overall, but when focusing on 5-to-4 decisions she wrote substantially more than expected, though Kennedy is up there with her and unsurprisingly the Chief Justice wrote even more.  Curiously, she wrote an average number of concurring opinions (Scalia had the most) in general, but when focusing on 5-to-4 decisions she demonstrably took the lead in this measure \cite[p.232]{SandraDayMostPowerful}.

In 2013, Enns and Wohlfarth extended the earlier work of Martin, Quinn, and Epstein \cite{MedianJustice-Martin-Quinn-Epstein} investigating median justices \cite{SwingJustice}.  They computed the ``5th vote'' in all cases from 1953 to 2009, meaning the 5th justice most likely to have joined the majority based on the locations of the one-dimensional Martin-Quinn scores.  Interestingly, they found that while the median justice is most frequently the 5th vote, many times a justice other than the median casts the decisive 5th vote \cite[p.1095]{SwingJustice}.  This finding leads them to posit that prior research on the swing vote is misleading: the concept of a swing vote should be case-specific, rather than term-specific, they argue.  Indeed, ``while the term-specific median justice casts the pivotal vote in more than half of all 5-to-4 decisions, more extreme justices account for 45\% of these pivotal swing votes." \cite[p.1103]{SwingJustice}.


\section{Methodology}

The first task we embark upon is to gain insight into the distribution of 5-to-4 decisions on the Court---or, put another way, insight into the coalitions of size five that have formed on the Court.  We rely on the widely-used Supreme Court Database \cite{SCDB} and declare a collection of five justices to be a \emph{voting coalition} if they provided the majority votes in a 5-to-4 decision.  In the 2009--2015 natural court, there were sixteen different voting coalitions (shown in Table \ref{tab:votingcoalitions}) out of the $\binom{9}{5} = 126$ total possible groupings of five justices.

\begin{table}[htp]
\caption{The sixteen voting coalitions occurring during the 2009--2015 natural court, listed alphabetically.  The breadth here hints at how central, yet perplexing, oddball coalitions are to Supreme Court jurisprudence.}
\begin{center}
\begin{tabular}{|c|c|c|c|c|c|}
\hline
(1) & Alito & Breyer & Ginsburg & Kagan & Kennedy\\
(2) & Alito & Breyer & Ginsburg & Roberts & Sotomayor\\
(3) & Alito & Breyer & Kennedy & Roberts & Sotomayor\\
(4) & Alito & Breyer & Kennedy & Roberts & Thomas\\
(5) & Alito & Breyer & Roberts & Scalia & Thomas\\
(6) & Alito & Kennedy & Roberts & Scalia & Thomas\\
(7) & Alito & Roberts & Scalia & Sotomayor & Thomas\\
(8) & Breyer & Ginsburg & Kagan & Kennedy & Sotomayor\\
(9) & Breyer & Ginsburg & Kagan & Roberts & Scalia\\
(10) & Breyer & Ginsburg & Kagan & Roberts & Sotomayor\\
(11) & Breyer & Ginsburg & Kagan & Sotomayor & Thomas\\
(12) & Breyer & Kagan & Kennedy & Roberts & Sotomayor\\
(13) & Ginsburg & Kagan & Kennedy & Roberts & Scalia\\
(14) & Ginsburg & Kagan & Kennedy & Scalia & Sotomayor\\
(15) & Ginsburg & Kagan & Scalia & Sotomayor & Thomas\\
(16) & Kagan & Kennedy & Scalia & Sotomayor & Thomas\\
\hline
\end{tabular}
\end{center}
\label{tab:votingcoalitions}
\end{table}

Only two of these are the ``obvious'' voting coalitions---the liberal coalition (8) and the conservative coalition (6)---so one naturally wonders where the fourteen remaining voting coalitions came from and how to interpret them, in a manner of speaking.  We first consider a combinatorial model in which justices shuffle along a one-dimensional ideal space; this is done primarily to encapsulate the prevailing political perspective of the Supreme Court.  Next, we introduce two geometric models relying on a two-dimensional MDS ideal space, one based on higher order Voronoi tessellations and one based on linear subdivisions of the ideal space.

After studying the majority coalitions involved in 5-to-4 decisions, the next topics we turn to are determining the decisive fifth vote in each 5-to-4 case and determining the median justice for each natural court (or range of terms within a natural court).  While both of these topics have previously been addressed in the literature from a statistical perspective, we introduce geometric methods based again on a two-dimensional MDS ideal space.  To estimate the fifth vote we find the majority justice most distant from the MDS focal point of the case, and to estimate the median justice we find the justice closest to the MDS focal point of all nine justices.  It is convenient to think of these focal points as exerting a ``circle of influence'' on the justices, as we shall see.


\subsection{Discrete disorder}\label{sec:disorder}

Edelman, Klein, and Lindquist introduced a quantity they call \emph{disorder} which measures how much a given coalition deviates from the coalitions one expects based on a one-dimensional voting model \cite[p.821]{EdelmanMeasuringDeviations}.  Their disorder scores are real numbers that depend on the coordinates of the nine justices in a one-dimensional ideal space.  We introduce here a discrete variant of their measure, so that we can work instead with integer-valued disorder scores; we focus on coalitions of size five, since that is our primary interest in this paper, but the definition readily extends to coalitions of any size.

We first use one-dimensional MDS to place the nine justices in a natural court along the real number line.\footnote{\label{foot:MDS}As noted earlier, this generally can be interpreted as a liberal-to-conservative scale.  Other methods of estimating one-dimensional ideal points are perfectly valid---we chose MDS to be consistent with the rest of the paper, where we do analyses based on two-dimensional MDS.  To compute the voting dissimilarity matrix that is fed into the classical MDS algorithm, we take the $ij$ entry to be one minus the following ratio: the number of cases in the Supreme Court Database for the given natural court where Justice $i$ and Justice $j$ have the same value for the ``majority'' variable, divided by the total number of cases in the natural court where all nine justices have a value recorded for this variable.}  Let us call the five leftmost justices on this scale and the five rightmost justices the two ``extreme blocs'' (which are typically liberal and conservative, respectively, though that interpretation is immaterial to the notion of disorder).  Given a particular 5-to-4 vote, the \emph{discrete disorder} is the minimum number of adjacent transpositions required to transform this voting coalition (i.e., the five majority justices in this case) into either of the two extreme blocs.  More formally, if we label the justices $1,2,\ldots,9$ based on their one-dimensional MDS ordering, then the discrete disorder of a voting coalition $J = \{j_1,j_2,j_3,j_4,j_5\}$, where $1 \le j_1 < j_2 < j_3 < j_4 < j_5 \le 9$, is 
\[disord(J) := \min\{\sum_{k=1}^5 j_k - k, ~ \sum_{k=1}^5 10 - k - j_{6-k}\}.\] In fact, we can use this formula to discuss the discrete disorder $disord(J)$ of any set $J$ of five justices, whether or not they formed a voting coalition.  

\begin{figure}
\begin{center}
\includegraphics[trim={0.8in 0.6in 0.4in 0.8in},clip,scale=0.9]{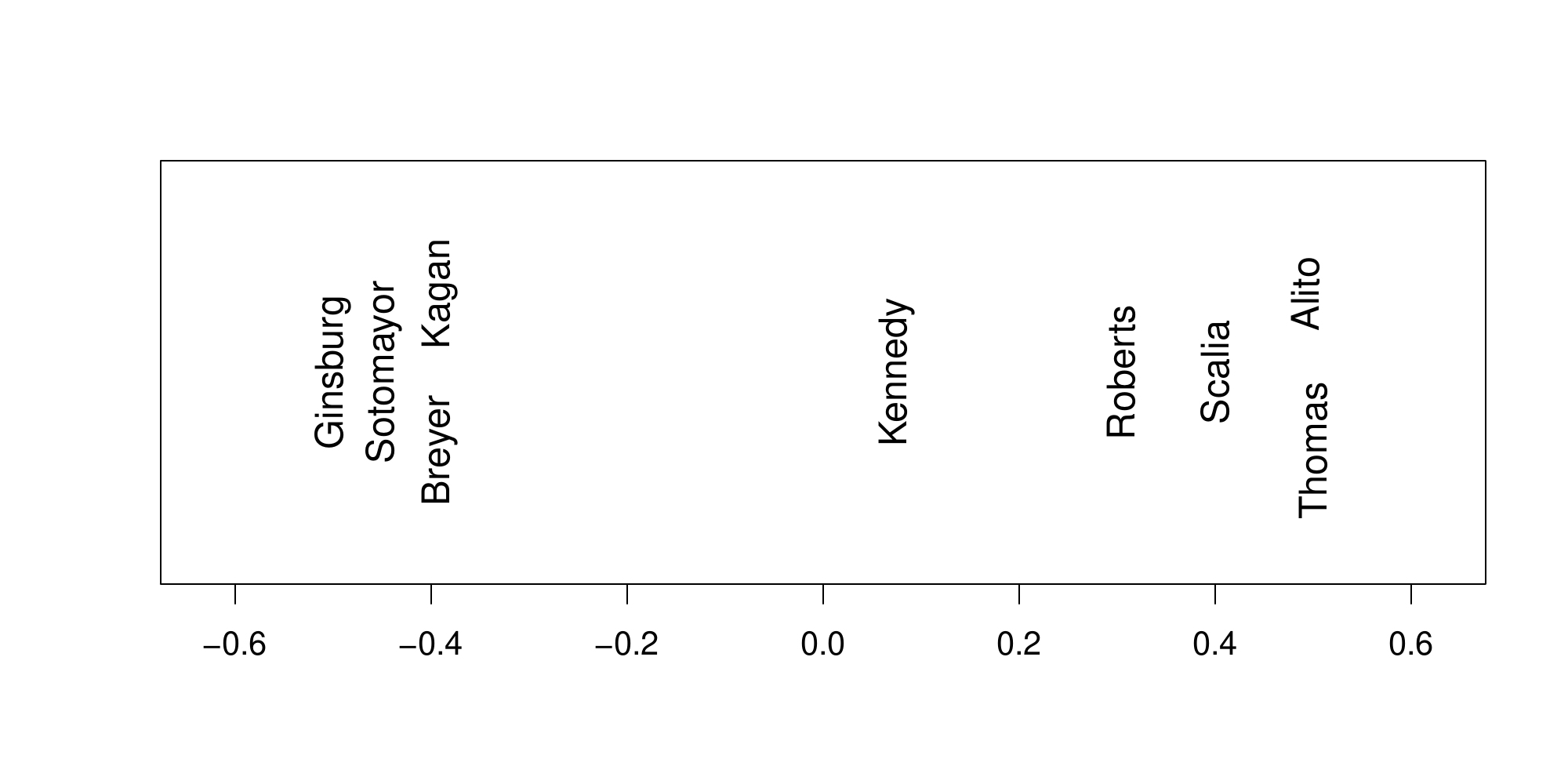}
\end{center}
\caption{The one-dimensional MDS ideal point locations of the nine justices on the 2009--2015 natural court.  Breyer is just slightly to the left of Kagan, and Alito is slightly to the left of Thomas.}
\label{fig:1mds}
\end{figure}

To illustrate this concept, let us consider again the 2009--2015 natural court.  Figure \ref{fig:1mds} shows the one-dimensional MDS ideal points for the nine justices, which orders the justices as follows:
\[\text{Ginsburg} <  \text{Sotomayor} < \text{Breyer} < \text{Kagan} < \text{Kennedy} < \text{Roberts} < \text{Scalia} < \text{Alito} < \text{Thomas}.\]
We see that the two extreme blocs, as expected, are the liberal and conservative coalitions (labelled (8) and (6), respectively, in Table \ref{tab:votingcoalitions}).  These extreme blocs are the only sets of five justices with discrete disorder score zero.  In Table \ref{tab:disordercoalitions} we list all the size five groups of justices with discrete disorder at most three and indicate  which of these occurs as a voting coalition.

\begin{table}[htp]
\caption{The sets of five justices with discrete disorder score at most three for the 2009--2015 natural court.  The number in the second column from the right is the discrete disorder score, and a check mark in the rightmost column indicates that this set of justices occurred during this natural court as a voting coalition.}
\begin{center}
\begin{tabular}{|c|c|c|c|c|c|c|}
\hline
Alito & Kennedy & Roberts & Scalia & Thomas & 0 & \checkmark \\
Breyer & Ginsburg & Kagan & Kennedy & Sotomayor & 0 & \checkmark \\
\hline
Alito & Kagan & Roberts & Scalia & Thomas & 1 & \\
Breyer & Ginsburg & Kagan & Roberts & Sotomayor & 1 & \checkmark \\
\hline
Alito & Breyer & Roberts & Scalia & Thomas & 2 & \checkmark \\
Alito & Kagan & Kennedy & Scalia & Thomas & 2 & \\
Breyer & Ginsburg & Kagan & Scalia & Sotomayor & 2 & \\
Breyer & Ginsburg & Kennedy & Roberts & Sotomayor & 2 & \\
\hline
Alito & Breyer & Kennedy & Scalia & Thomas & 3 & \\
Alito & Breyer & Ginsburg & Kagan & Sotomayor & 3 & \\
Alito & Kagan & Kennedy & Roberts & Thomas & 3 & \\
Alito & Roberts & Scalia & Sotomayor & Thomas & 3 & \checkmark \\
Breyer & Ginsburg & Kennedy & Scalia & Sotomayor & 3 & \\
Ginsburg & Kagan & Kennedy & Roberts & Sotomayor & 3 & \\
\hline
\end{tabular}
\end{center}
\label{tab:disordercoalitions}
\end{table}

Note that many groups of five justices have a low disorder score and yet do not appear in any of the 349 cases tallied\footnote{Recall that we only consider cases in the Supreme Court Database where all nine justices have a value recorded for the ``majority'' variable.} during this natural court.  On the other hand, the discrete disorder scores for the sixteen voting coalitions, in increasing order, are: \[0, 0, 1, 2, 3, 4, 4, 4, 5, 5, 6, 6, 8, 8, 8, 9.\] This illustrates the challenge with viewing oddball coalitions simply as a matter of disorder: many low discrete disorder score groupings never occur, and many groupings that do occur (i.e., voting coalitions) have strikingly high discrete disorder scores.  For instance, the highest discrete disorder score among the sixteen voting coalitions seen during this natural court is nine,\footnote{This voting coalition occurred exactly once during this natural court, in 2015 when Alito, Breyer, Kennedy, Roberts, and Sotomayor formed the majority in \emph{Comptroller of the Treasury of Maryland v.\ Wynne}.} yet the maximum theoretically possible discrete disorder score is only ten, so this coalition is almost as disordered as one could ever be.  Moreover, of the 126 possible groupings, 114 of these have discrete disorder score at most nine---so discrete disorder leaves much to be desired in terms of explaining which voting coalitions one expects to see on the Court.


\subsection{Voronoi coalitions}

Voronoi diagrams first started gaining prominence in the 19th century, in both pure mathematics and epidemiology, and have since found widespread use in a multitude of settings \cite{Voronoi}.  A two-dimensional Voronoi diagram is a particular way of partitioning the plane into convex regions: given a finite collection of ``seed'' points, there is a convex region around each seed point consisting of all points in the plane closer to the given seed point than to all the other seed points.  A related construction is the higher order Voronoi diagram: the $k$-th order Voronoi diagram partitions the plane into convex regions consisting of all points in the plane that have the same set of $k$ nearest seed points \cite{HigherVoronoi}.  

To study 5-to-4 vote patterns on a fixed natural court, we consider the two-dimensional MDS locations of the nine justices  (see Footnote \ref{foot:MDS} above for the technical details of what this means) and take these as our seed points and then construct the corresponding 5th order Voronoi diagram.  Thus each convex region corresponds to a collection of five justices with some ideological focal point in common---meaning that there is a point in our MDS plane such that these five justices are precisely the five closest justices to that point.  For a given natural court, we compute all the sets of five justices that arise this way, let us call them \emph{Voronoi coalitions}, and our primary question is how these compare to the voting coalitions on that natural court.  Notably, if an oddball voting coalition happens to be a Voronoi coalition, then in some sense this ``explains'' the oddball coalition since it reveals a geometric commonality uniting the five justices.  

\begin{table}[htp]
\caption{The twenty Voronoi coalitions for the 2009--2015 natural court.  The numerical column displays the discrete disorder score (\S\ref{sec:disorder}) and a checkmark indicates the justices formed a voting coalition on this natural court (cf., Table \ref{tab:votingcoalitions}).}
\begin{center}
\begin{tabular}{|c|c|c|c|c|c|c|}
\hline
Alito & Breyer & Kagan & Kennedy & Roberts &  9 &  \\
Alito & Breyer & Kennedy & Roberts & Scalia & 6 & \\
Alito & Breyer & Kennedy & Roberts & Sotomayor & 9 & \checkmark \\
Alito & Breyer & Kennedy & Roberts & Thomas & 4 & \checkmark \\
Alito & Kagan & Kennedy & Roberts & Scalia & 5 & \\
Alito & Kagan & Roberts & Scalia & Thomas & 1 & \\
Alito & Kennedy & Roberts & Scalia & Thomas & 0 & \checkmark \\
Breyer & Ginsburg & Kagan & Kennedy & Sotomayor & 0 & \checkmark \\
Breyer & Ginsburg & Kagan & Scalia & Sotomayor & 2 & \\
Breyer & Ginsburg & Kennedy & Roberts & Sotomayor & 2 & \\
Breyer & Kagan & Kennedy & Roberts & Scalia & 10 & \\
Breyer & Kagan & Kennedy & Roberts & Sotomayor & 5 & \checkmark \\
Ginsburg & Kagan & Kennedy & Roberts & Scalia & 8 & \checkmark \\
Ginsburg & Kagan & Kennedy & Roberts & Sotomayor & 3 & \\
Ginsburg & Kagan & Kennedy & Scalia & Sotomayor & 4 & \checkmark \\
Ginsburg & Kagan & Kennedy & Scalia & Thomas & 9 & \\
Ginsburg & Kagan & Roberts & Scalia & Thomas & 8 & \\
Ginsburg & Kagan & Scalia & Sotomayor & Thomas & 8 & \checkmark \\
Kagan & Kennedy & Roberts & Scalia & Sotomayor & 9 & \\
Kagan & Kennedy & Roberts & Scalia & Thomas & 4 &\\
\hline
\end{tabular}
\end{center}
\label{tab:Voronoicoalitions}
\end{table}

\begin{figure}
\begin{center}
\includegraphics[scale=0.4]{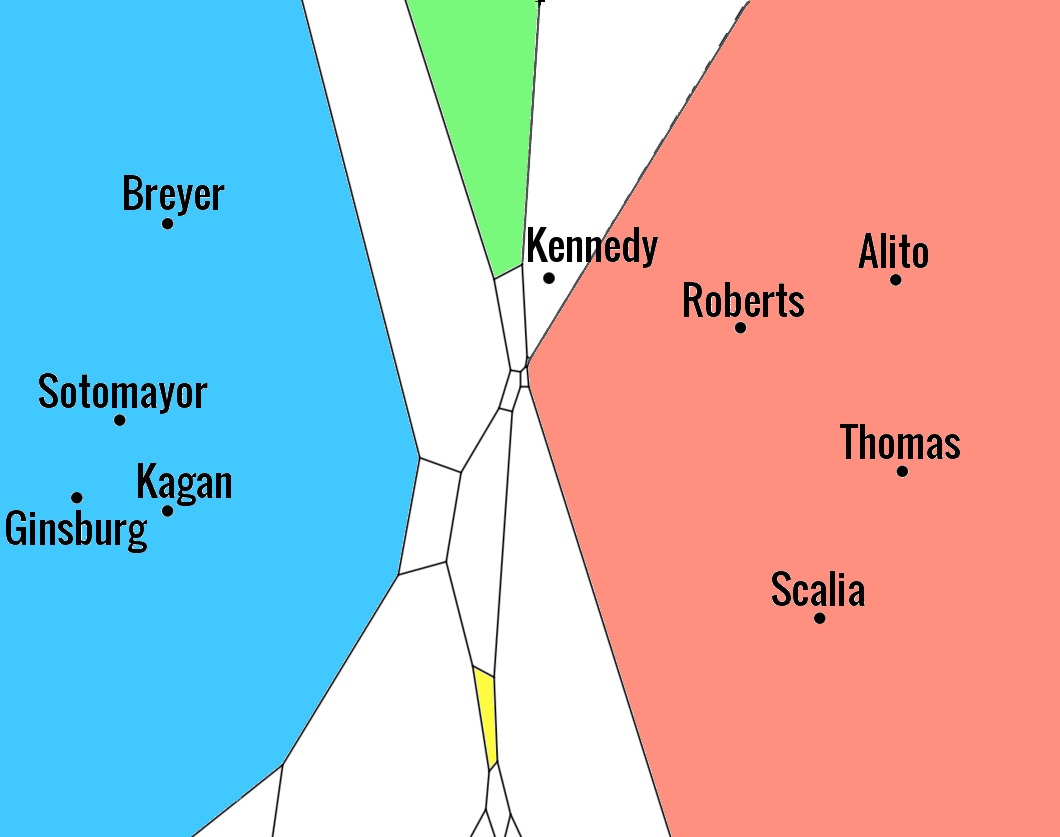}
\end{center}
\caption{The 5th order Voronoi diagram for the two-dimensional MDS locations of the nine justices in the 2009--2015 natural court.  The blue region (left) consists of all points whose five nearest justices are the liberal coalition of Breyer, Ginsburg, Kagan, Kennedy, and Sotomayor.  The red region (right) corresponds to the conservative coalition: Alito, Kennedy, Roberts, Scalia, and Thomas.  The green region (top) corresponds to the voting coalition with the highest discrete disorder score: Alito, Breyer, Kennedy, Roberts, and Sotomayor.  The yellow region (bottom) corresponds to another interesting oddball voting coalition: Ginsburg, Kagan, Kennedy, Roberts, and Scalia.}
\label{fig:Vor5cols}
\end{figure}

The Voronoi coalitions for the 2009--2015 natural court listed in Table \ref{tab:Voronoicoalitions} show that 40\% of the Voronoi coalitions are voting coalitions, and 50\% of the voting coalitions are Voronoi coalitions.  Thus, roughly half of the 5-to-4 cases in this natural court comport with our Voronoi model of Supreme Court voting---which is quite remarkable given how many oddball, highly disordered coalitions arise here and given the limited progress in the literature so far in explaining them conceptually.  In particular, the most disordered voting coalition from this natural court (the Alito-Breyer-Kennedy-Roberts-Sotomayor majority in the 5-to-4 case \emph{Comptroller of the Treasury of Maryland v.\ Wynne}) quite strikingly is a Voronoi coalition (see Figure \ref{fig:Vor5cols})---so the case that a traditional one-dimensional, liberal-to-conservative model struggles the most to explain fits perfectly in our two-dimensional Voronoi spatial model.

\subsection{Halving lines}

Edelman describes spatial voting models as an ``attempt to identify each Justice with an ideal point in a policy space and then model issues as cutting planes in such a way that the set of Justices on one side of the plane vote one way and those on the other vote the opposite'' \cite[p.10]{DimSupCourtEdelman}.  However, we have been unable to find in the literature on Supreme Court rulings a systematic implementation and analysis of this concept for dimensions greater than one, so we embark upon that here---in the specific context of 5-to-4 decisions.  Namely, we fix a natural court and consider the nine justices in planar coordinates given by two-dimensional MDS and then declare any five justices to be a \emph{half-plane coalition} if there exists a line dividing the plane such that the five justices lie on one side and the remaining four justices lie on the other side.  In the mathematics literature, specifically in discrete geometry, these half-plane coalitions are called ``$k$-sets'' for $k=5$ \cite{k-sets}.  This provides another two-dimensional spatial model of Supreme Court voting---the key here is that, in contrast to the work of Fischman and Jacobi \cite{2ndDimension} where only vertical and horizontal lines are considered, lines of arbitrary slope are allowed and thereby split the justices along a combination of the traits characterized by each dimension.  As with the Voronoi coalitions, we are interested in how these half-plane coalitions compare with, and help explain, the voting coalitions.

\begin{figure}
\begin{center}
\includegraphics[scale=0.3]{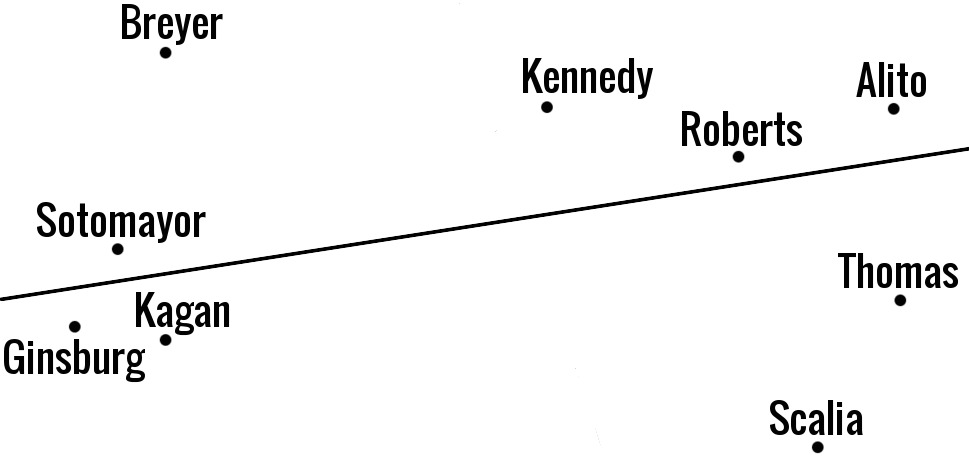}
\end{center}
\caption{Illustration of a line dividing the 2009--2015 natural court MDS plane into two halves, showing that the oddball voting coalition of Alito, Breyer, Kennedy, Roberts, and Sotomayor is a half-plane coalition.}
\label{fig:halfplane}
\end{figure}

\begin{table}[htp]
\caption{The ten half-plane coalitions for the 2009--2015 natural court.  The numerical column displays the discrete disorder score (\S\ref{sec:disorder}) and a checkmark indicates the justices formed a voting coalition on this natural court.}
\begin{center}
\begin{tabular}{|c|c|c|c|c|c|c|}
\hline
Alito & Breyer & Ginsburg & Kennedy & Sotomayor & 4 &  \\
Alito & Breyer & Kennedy & Roberts & Sotomayor & 9 & \checkmark \\
Alito & Breyer & Kennedy & Roberts & Thomas & 4 & \checkmark \\
Alito & Kagan & Roberts & Scalia & Thomas & 1 &  \\
Alito & Kennedy & Roberts & Scalia & Thomas & 0 & \checkmark \\
Breyer & Ginsburg & Kagan & Kennedy & Sotomayor & 0 & \checkmark \\
Breyer & Ginsburg & Kagan & Scalia & Sotomayor & 2 &  \\
Breyer & Ginsburg & Kennedy & Roberts & Sotomayor & 2 &  \\
Ginsburg & Kagan & Roberts & Scalia & Thomas & 8 &  \\
Ginsburg & Kagan & Scalia & Sotomayor & Thomas & 8 & \checkmark \\
\hline
\end{tabular}
\end{center}
\label{tab:halfplanecoalitions}
\end{table}
For the 2009--2015 natural court (see Table \ref{tab:halfplanecoalitions}), 50\% of the half-plane coalitions are also voting coalitions and 31.25\% of the voting coalitions are half-plane coalitions; this model also explains some highly disordered odd-ball coalitions (for an example, see Figure \ref{fig:halfplane}).  Curiously, on this natural court every group of five justices that is simultaneously a voting coalition and a half-plane coalition is also a Voronoi coalition.

\subsection{Determining the fifth vote}\label{sec:fifth}

Perhaps the most important question one can ask about a 5-to-4 vote on the Supreme Court is which of the five justices in the majority came closest to voting with the four-justice minority, thereby flipping the ruling to the opposite outcome.  While such information cannot truly be ascertained, we nonetheless use our two-dimensional spatial model to propose two methods for ranking the majority votes so that we can label one justice as having provided the crucial ``fifth'' vote in a 5-to-4 decision (see \cite{SwingJustice} for another approach).

For a particular natural court, we first consider the two-dimensional MDS layout of the nine justices.  Next, for each 5-to-4 vote we consider the five justices in the majority and take the mean of their MDS coordinates; this yields what can be interpreted as the focal point of the majority perspective of the case.  Then, we order the majority justices by their distance from this focal point---from closest to furthest---and declare the fifth one to be the one who cast the decisive \emph{fifth vote}.  In a distinct but related manner, we also take the four minority justices, compute their mean coordinates to identify the focal point of the minority perspective, then order the majority justices based on their distance to this focal point---this time, from furthest to closest---and declare the fifth one to have cast the decisive fifth vote.  The idea here is that in the first method we have identified the majority justice least strongly allied with the majority outlook, and in the second method we found the majority justice most sympathetic to the minority outlook.  

In some situations these two methods yield the same fifth vote, but in others they do not.  For the voting coalitions listed in Table \ref{tab:votingcoalitions}, nine out of sixteen times the two methods agree.  The liberal coalition and the conservative coalition (numbers (8) and (6), respectively, in that table) both have Kennedy as the fifth vote for both methods, which matches the general view of Kennedy as the predominate swing vote on that Court.  The voting coalition numbered (10) in Table \ref{tab:votingcoalitions}, which occurred for instance in the famous Affordable Care Act case \emph{National Federation of Independent Businesses v. Sebelius}, has Roberts labelled as the fifth vote by both methods; this supports the standard view that in this case Roberts unexpectedly cast the decisive vote supporting a generally liberal cause.  Another interesting example is the voting coalition numbered (14) in Table \ref{tab:votingcoalitions}, which occurred in a variety of cases; here the first method (based on proximity to majority focal point) finds Scalia as the fifth vote whereas the second method (based on proximity to the minority focal point) finds Kennedy as the fifth vote. 

\begin{figure}
\begin{center}
\includegraphics[trim={0.8in 2.4in 0.4in 2.2in},clip,scale=0.88]{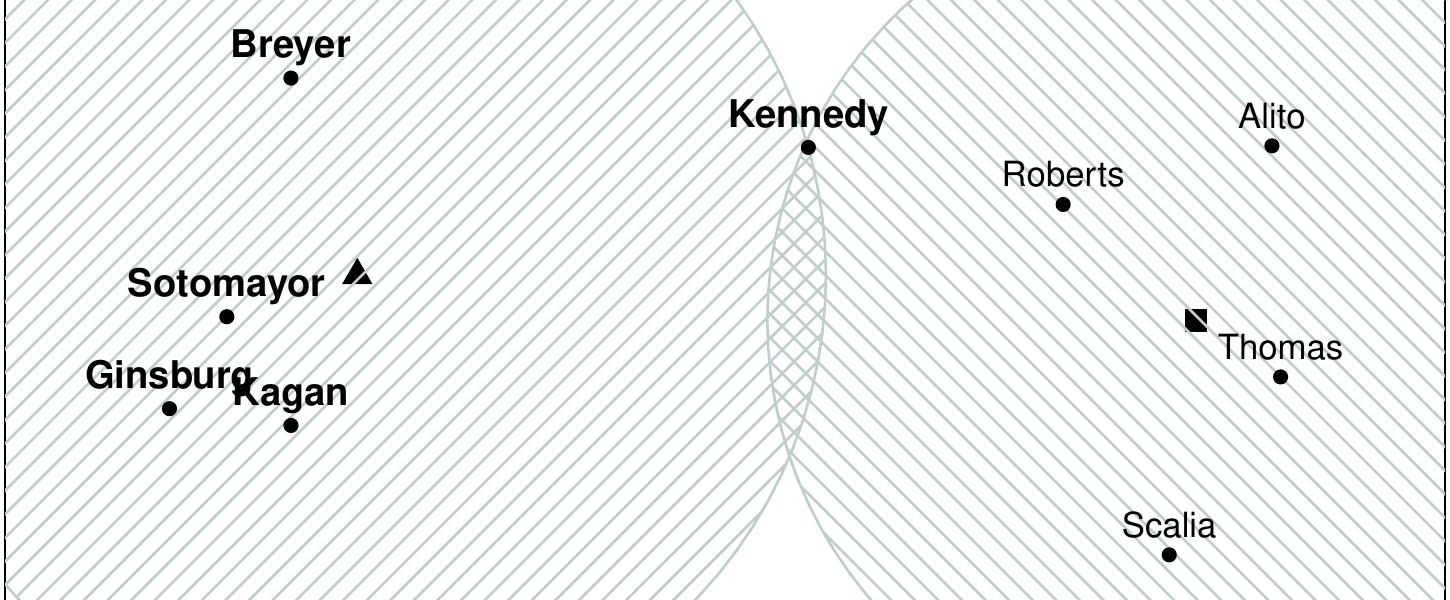}\\
(a)\\
\includegraphics[trim={0.8in 2.4in 0.4in 2.2in},clip,scale=0.88]{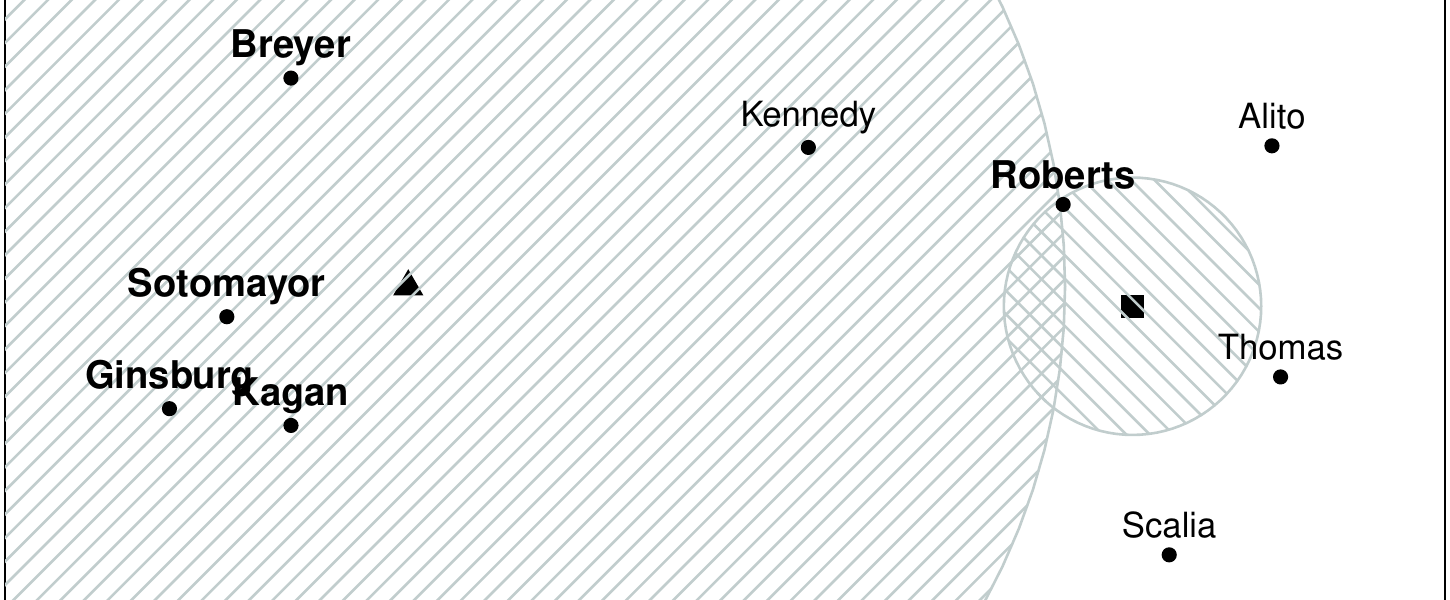}\\
(b)\\
\includegraphics[trim={0.8in 2.4in 0.4in 2.2in},clip,scale=0.88]{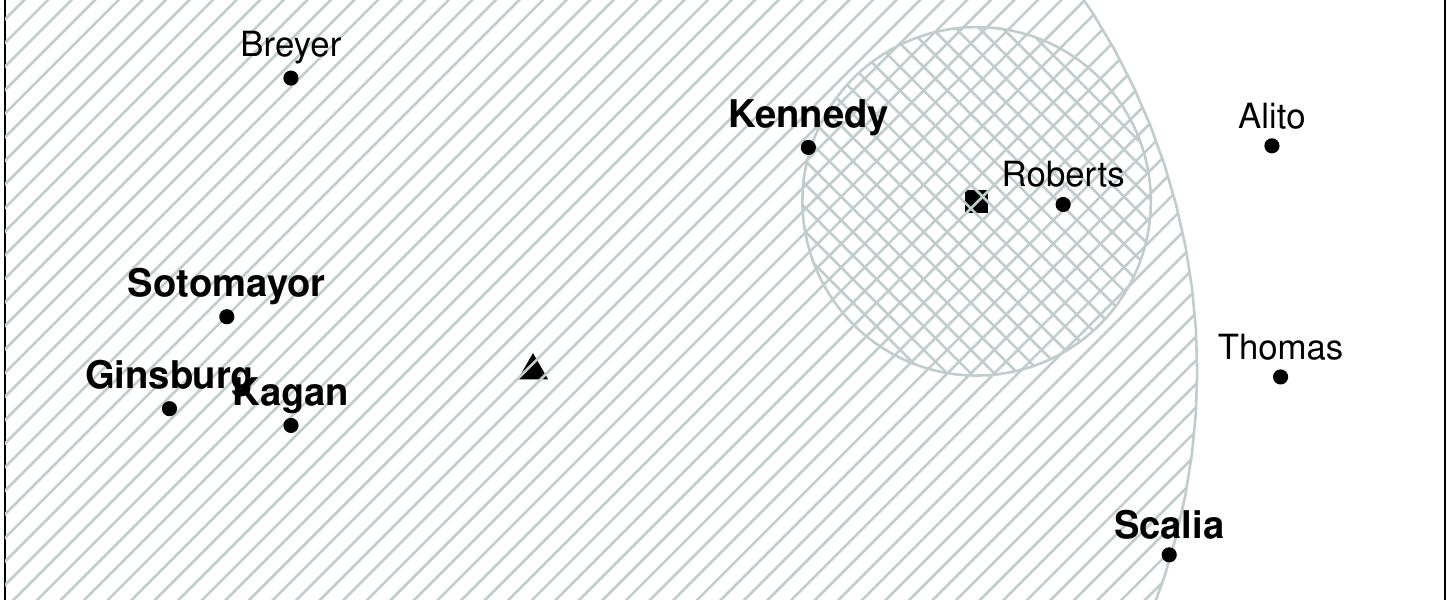}\\
(c)
\end{center}
\caption{Circles of influence illustrating the decisive fifth vote for three different voting coalitions.  In all three cases the names of the majority votes are in bold and their focal point is indicated by a triangle {\small $\blacktriangle$} whereas the minority focal point is a square {\tiny $\blacksquare$}.  (a) The majority is the liberal coalition and both spatial methods find Kennedy as the fifth vote.  (b) This voting coalition occurred, among other instances, in the 2012 ruling supporting the Affordable Care Act and both spatial methods find Roberts as the fifth vote.  (c) The two spatial methods disagree on this voting coalition: Scalia is the fifth vote when using the majority focal point, whereas Kennedy is the fifth vote when using the minority focal point.}
\label{fig:fifth}
\end{figure}

It is convenient to think of these ideological focal points as exerting circles of influence, with the justices closest to the center of the circle most strongly aligned with the corresponding outlook on the case.  In a 5-to-4 decision the majority focal point reaches outward and collects five justices, the last one being the least steadfast in his or her vote; the minority focal point reaches outward and the first justice in the majority it reaches is the one most tempted to defect.  This idea is illustrated in Figure \ref{fig:fifth} with the voting coalitions discussed in the preceding paragraph.

\subsection{The mean justice}\label{sec:meanjustice}

We now turn from the case-specific swing vote to the term-specific and court-specific swing vote.  Given a natural court or collection of cases within a natural court, we first compute the two-dimensional MDS layout of the justices corresponding to the voting similarity matrix for these cases.  Next, we compute the mean of the two coordinates of all nine justices to find the focal point of the entire Court.  We then label the justice whose Euclidean distance to this focal point is minimal the \emph{mean justice}.  Since we are dealing with two dimensions rather than one we must use a mean rather than a median---so our choice of terminology here is to suggest a similarity to, but distinction from, the notion of ``median justice'' that appears frequently in the literature.

For the 2009--2015 natural court, the mean justice is Kennedy; this supports the common view of him as the predominate swing vote on this Court.  When computing MDS coordinates for each term separately, we find that Kennedy was the mean justice for each term within this natural court except for 2013, where the distinction goes to Roberts, and 2015, where Breyer becomes the mean justice.  Since Roberts was the mean justice for the 2013 term, this perhaps helps explain why he was in a position to cast the decisive fifth vote in the 2012 case \emph{National Federation of Independent Businesses v. Sebelius}---though this does not explain the particular direction of his vote, which is admittedly the far more surprising and significant aspect of this case.


\section{Results}

We begin with a broad view of our spatial models and then focus on a specific case to illustrate the insight that can be gleaned from our geometric perspective.

\begin{table}[htp]
\caption{Overview of our spatial methods applied to the natural courts in the Supreme Court Database \cite{SCDB}.  ``Voting coalitions'' is the number of distinct voting coalitions (i.e., groups of five justices forming the majority in a 5-to-4 case) in the specified natural court.  ``Max disorder'' is the highest discrete disorder score among the voting coalitions.  ``Voronoi accuracy'' lists the percentage of voting coalitions that are also Voronoi coalitions, then the percentage of Voronoi coalitions that are also voting coalitions.  ``Half-plane accuracy'' lists the percentage of voting coalitions that are also half-plane coalitions, then the percentage of half-plane coalitions that are also voting coalitions. ``Mean justice'' is defined in \S\ref{sec:meanjustice}.}
\begin{center}
\begin{tabular}{|c|c|c|c|c|c|}
\hline
Natural & Voting & Max & Voronoi & Half-plane & Mean\\  
court & coalitions & disorder & accuracy & Half-plane & justice\\  
\hline
1946--1948 & 28 & 10 & 35.7\%, 50\% & 25\%, 70\% & Vinson \\
1949--1952 & 26 & 9 & 30.8\%, 50\% & 23.1\%, 42.9\% & Burton \\
1953--1953 & 8 & 10 & 37.5\%, 15.8\% & 37.5\%, 27.3\% & Clark \\
1954--1956 & 8 & 6 & 75\%, 28.6\% & 62.5\%, 55.6\% & Clark \\
1956--1956 & 5 & 8 & 80\%, 21.1\% & 80\%, 36.4\% & Brennan \\
1956--1958 & 10 & 9 & 60\%, 31.6\% & 40\%, 40\% & Brennan \\
1958--1961 & 18 & 9 & 44.4\%, 40\% & 33.3\%, 60\% & Brennan \\
1962--1964 & 18 & 9 & 33.3\%, 33.3\% & 33.3\%, 40\% & White \\
1965--1966 & 12 & 9 & 58.3\%, 38.9\% & 58.3\%, 58.3\% & White \\
1967--1968 & 6 & 10 & 50\%, 15.8\% & 16.7, 8.3\% & Marshall \\
1969--1970 & 15 & 7 & 33.3\%, 26.3\% & 33.3\%, 50\% & White \\
1971--1975 & 23 & 10 & 43.5\%, 50\% & 39.1\%, 90\% & Stewart \\
1975--1980 & 33 & 10 & 33.3\%, 57.9\% & 30.3\%, 90.9\% & Powell \\
1981--1985 & 33 & 10 & 33.3\%, 57.9\% & 27.3\%, 81.8\% & White \\
1986--1986 & 13 & 8 & 38.5\%, 26.3\% & 38.5\%, 45.5\% & Powell \\
1987--1989 & 19 & 8 & 42.1\%, 44.4\% & 31.6\%, 60\% & White \\
1990--1990 & 15 & 10 & 60\%, 47.4\% & 60\%, 75\% & Souter \\
1991--1992 & 20 & 9 & 40\%, 44.4\% & 40\%, 66.7\% & Souter \\
1993--1993 & 6 & 9 & 66.7\%, 22.2\% & 66.7\%, 36.4\% & Kennedy \\
1994--2004 & 38 & 10 & 34.2\%, 72.2\% & 28.9\%, 100\% & Kennedy \\
2005--2005 & 2 & 0 & 100\%, 9.5\% & 100\%, 16.7\% & O'Connor \\
2005--2008 & 14 & 10 & 42.9\%, 31.6\% & 42.9\%, 60\% & Kennedy \\
2009--2009 & 8 & 10 & 50\%, 22.2\% & 50\%, 40\% & Kennedy \\
2009--2015 & 16 & 9 & 50\%, 40\% & 31.2\%, 50\% & Kennedy \\
2016--2016 & 2 & 0 & 100\%, 12.5\% & 100\%, 14.3\% & Kennedy \\
\hline
\end{tabular}
\end{center}
\label{tab:rates}
\end{table}

\begin{figure}
\begin{center}
\includegraphics[scale=0.83]{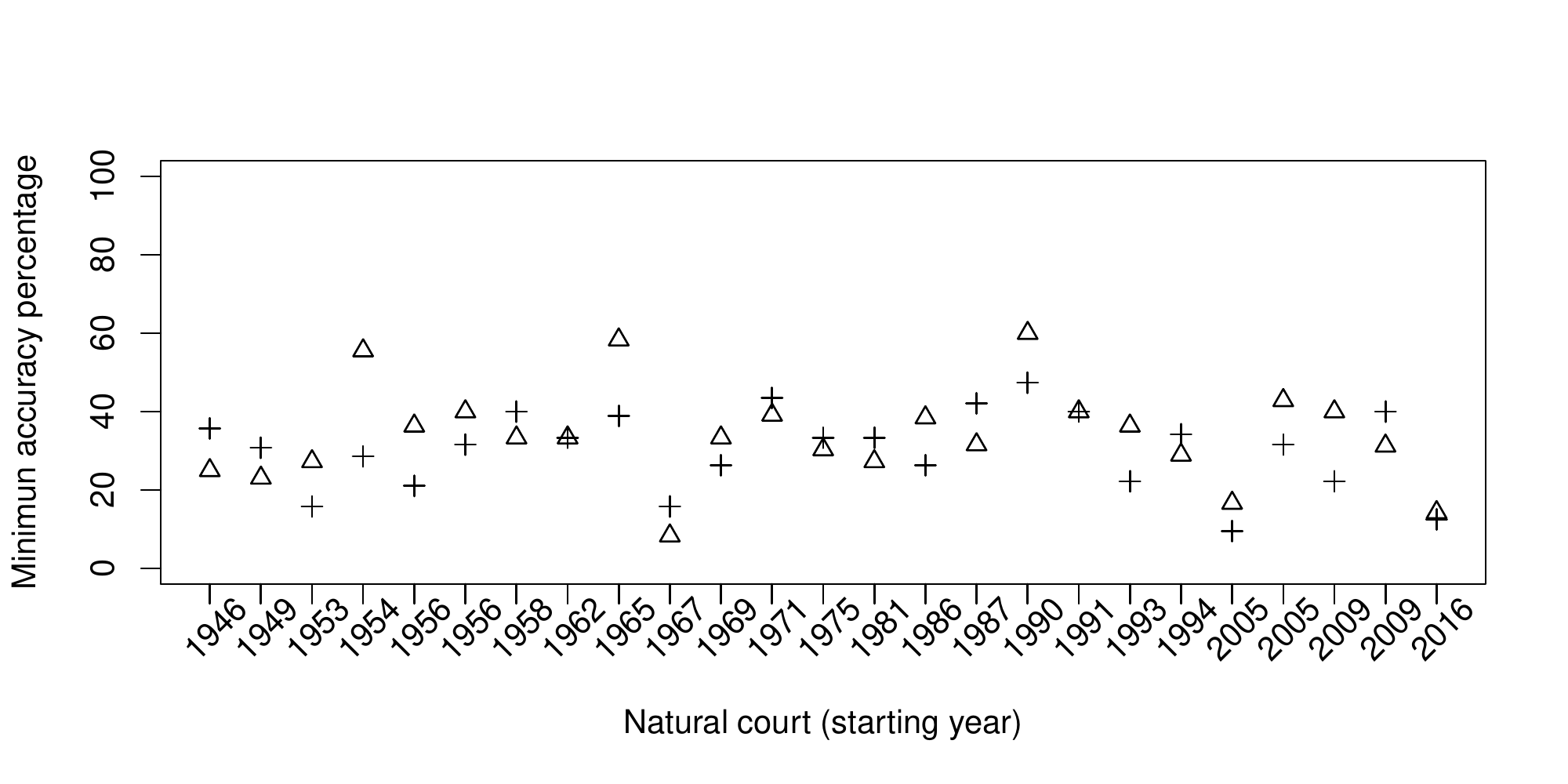}\\
\end{center}
\caption{The minimum of the first and second accuracy rates (``explained'' percentage and ``efficiency'' percentage) for each of the natural courts in the Supreme Court Database; plus $+$ denotes the Voronoi model, triangle {\footnotesize $\triangle$} denotes the half-plane model.}
\label{fig:minrates}
\end{figure}

For each of the two accuracy columns listed in Table \ref{tab:rates}, the first number should be thought of as the percentage of voting coalitions ``explained'' by the specified spatial model, while the second number measures how ``efficiently'' the model explains these voting coalitions.  A model that simply lists all possible sets of five justices would always have a perfect 100\% for its first accuracy score but its second accuracy score would be extremely low.  On the other hand, a model that only predicts two coalitions, the liberal coalition and the conservative coalition, would have a nearly perfect second accuracy score---but aside from the brief natural courts occurring during the 2005 and 2016 terms, such a model would generally have a very low first accuracy score.  These considerations show that it is not difficult to construct a model where the average of the two accuracy scores is at least 50\% for each natural court; what is more challenging, and significant, is constructing models where the \emph{minimum} of the two accuracy scores is as high as possible for each natural court---see Figure \ref{fig:minrates} for this measure of our two spatial models.

We find some rather striking results in Table \ref{tab:rates}.  For instance, the long 1994--2004 natural court saw a whopping thirty-eight different voting coalitions, and the half-plane model only found eleven coalitions but every single one it found is a true voting coalition.  The discrete disorder scores for these eleven coalitions range from zero to nine.  The coalition with discrete disorder nine is the majority of Breyer, Ginsburg, Kennedy, O'Connor, and Rehnquist, and a minority of Scalia, Souter, Stevens, and Thomas, which occurred in precisely one ruling, namely, the 2005 case \emph{United States v. Booker} on U.S. Sentencing Guidelines.  The half-plane model also found eleven coalitions on the 1975--1980 natural court, of which ten were among the thirty-three actual voting coalitions, and it found ten coalitions on the 1971--1975 natural court, of which nine were among the twenty-three actual voting coalitions.  Both the half-plane model and the Voronoi model found four out of the five voting coalitions on the 1956 natural court; both methods missed the majority of Black, Burton, Harlan, Douglas, and Warren, versus the minority of Brennan, Clark, Frankfurter, and Reed, which occurred in the 1956 case \emph{Massachusetts Bonding \& Ins. Company v.\ United States}.    

The first accuracy scores for the Voronoi model tend to be slightly higher than those of the half-plane model, while the second accuracy scores tend to be somewhat lower.  Indeed, the mean first accuracy rates are 50.9\% for Voronoi and 45.2\% for half-plane, while the mean second accuracy rates are 35.6\% for Voronoi and 52.6\% for half-plane.  A particularly strong year for both the Voronoi and half-plane models is the rather wild 1990 natural court, which saw fifteen voting coalitions in a single term.

Our mean justice column in Table \ref{tab:rates} supports the general view of Kennedy as a swing vote---or at least as a justice central to the Court's behavior---for a lengthy stretch of years.  Powell and O'Connor, both considered swing votes by various authors (cf., \S\ref{sec:backgroundswing}) also show up as the mean justice.  It is interesting to compare the natural court mean justices with the term-specific mean justices (see Table \ref{tab:mJ}).

\begin{table}[htp]
\caption{The term-specific mean justices for the past forty years.  Bold names are also the mean justice for the corresponding natural court.}
\begin{center}
\begin{tabular}{|c|c|c|c|c|c|c|c|}
\hline
Term & Mean justice & Term & Mean justice & Term & Mean justice & Term & Mean justice\\  
\hline
1977 & Stewart & 1987 & \textbf{White} & 1997 & \textbf{Kennedy} & 2007 & Roberts\\
1978 & Blackmun & 1988 & \textbf{White} & 1998 & \textbf{Kennedy} & 2008 & \textbf{Kennedy}\\
1979 & White & 1989 & \textbf{White} & 1999 & O'Connor & 2009 & \textbf{Kennedy}\\
1980 & White & 1990 & \textbf{Souter} & 2000 & O'Connor & 2010 & \textbf{Kennedy}\\
1981 & Blackmun & 1991 & Kennedy & 2001 & \textbf{Kennedy} & 2011 & \textbf{Kennedy}\\
1982 & Burger & 1992 & Kennedy & 2002 & \textbf{Kennedy} & 2012 & \textbf{Kennedy}\\
1983 & \textbf{White} & 1993 & \textbf{Kennedy} & 2003 & O'Connor & 2013 & Roberts\\
1984 & Powell & 1994 & \textbf{Kennedy} & 2004 & \textbf{Kennedy} & 2014 & \textbf{Kennedy}\\
1985 & Powell & 1995 & \textbf{Kennedy} & 2005 & \textbf{O'Connor} & 2015 & Breyer\\
1986 & \textbf{Powell} & 1996 & O'Connor & 2006 & \textbf{Kennedy} & 2016 & \textbf{Kennedy}\\
\hline
\end{tabular}
\end{center}
\label{tab:mJ}
\end{table}

While Epstein and Jacobi \cite{SuperMedians} introduce the notion of a ``super median justice,'' here we might estimate the strength of a mean justice by noting how often the term-specific and natural court-specific mean justices coincide.  We see in Table \ref{tab:mJ} that Kennedy certainly qualifies as a strong mean justice, and arguably White does as well.

\subsection{Illustration of methods on a single case}

Let us consider the 2001 case \emph{Rogers v. Tennessee}, which was also used by Edelman, Klein, and Lindquist to motivate their study of disordered Supreme Court votes \cite{EdelmanMeasuringDeviations}. We quote from Oyez:
\begin{quote}
Wilbert K. Rogers was convicted in Tennessee of second degree murder. The victim, James Bowdery, died 15 months after Rogers stabbed him. On appeal, Rogers argued that the Tennessee common law ``year and a day rule,'' under which no defendant could be convicted of murder unless his victim died by the defendant's act within a year and a day of the act, persisted and precluded his conviction. The Tennessee Court of Criminal Appeals affirmed the conviction. In affirming, the Tennessee Supreme Court ultimately abolished the rule and upheld Rogers' conviction. \cite{OyezRogers}
\end{quote}
Edelman et al.\ nicely frame the oddity of the oddball coalition that decided this case:
\begin{quote}
The Rogers case raised the important civil liberties question of whether judicially created criminal rules may be applied retroactively under the Due Process Clause. On the U.S. Supreme Court, one might expect such a case to trigger a predictable lineup of the Justices' votes, with conservatives adopting the ``law and order'' interpretation and liberals the prodefendant's rights approach. Yet the votes did not fall out this way. Instead, O'Connor's opinion of the Court, holding that Rogers's due process rights were not violated by the retroactive application of the rule was joined not just by conservatives Rehnquist and Kennedy, but also by moderate liberals Souter and Ginsburg. The most liberal member of the Court, Stevens, was joined in dissent by the moderate Breyer and the strongly conservative Thomas and Scalia.
\cite[p.820]{EdelmanMeasuringDeviations}
\end{quote}

\begin{figure}
\begin{center}
\includegraphics[trim={0.9in 2.1in 0.5in 1.8in},clip,scale=1.0]{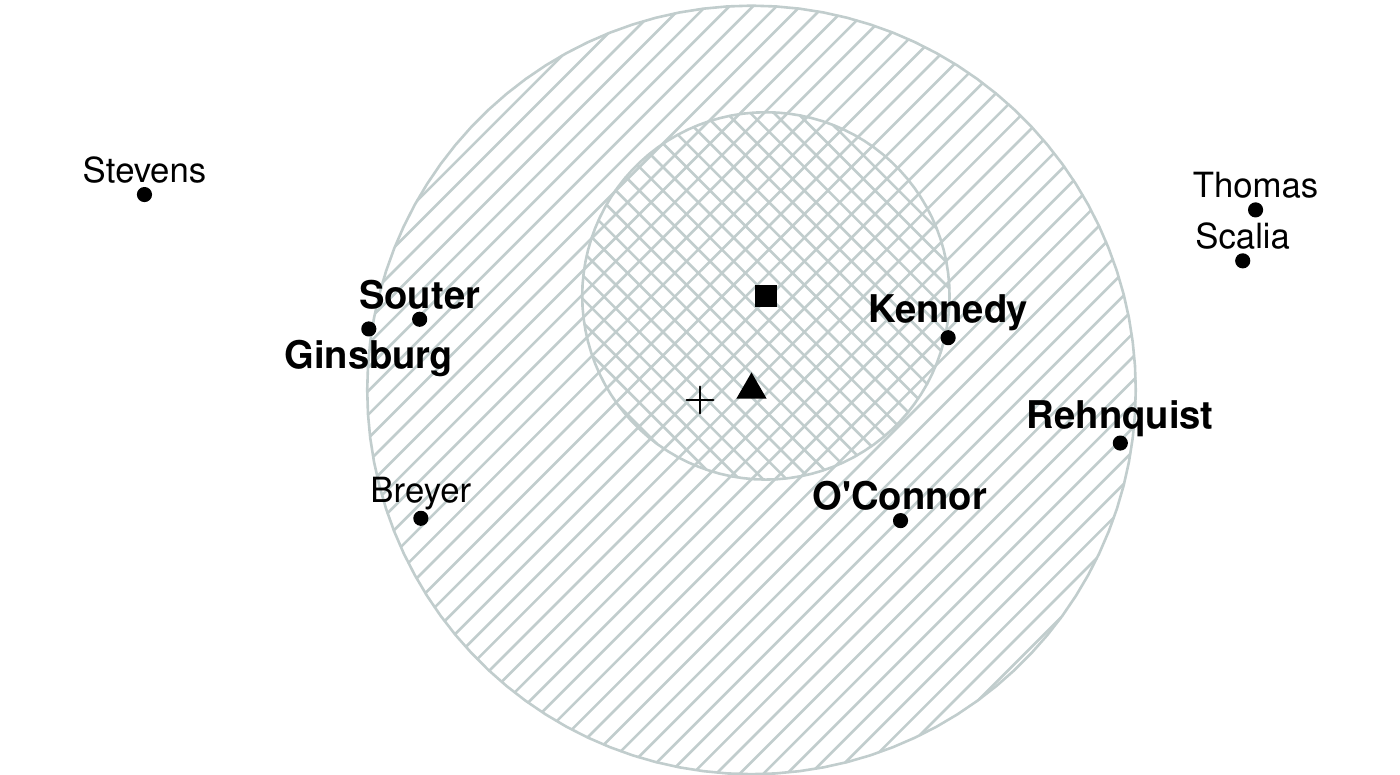}\\
\end{center}
\caption{The two-dimensional MDS layout of the justices on the 1994--2004 natural court, with the majority votes in the 2001 case \emph{Rogers v. Tennessee} in bold.  The majority focal point is marked with a triangle {\small $\blacktriangle$}, the minority focal point is marked with a square {\tiny $\blacksquare$}, and a plus $+$ marks the center of the entire Court.  The inner circle reaches out from the minority focal point and touches the closest majority justice, Kennedy, while the outer circle reaches out from the majority focal point to the fifth majority vote, namely Ginsburg.}
\label{fig:fifthcase}
\end{figure}

To begin our spatial analysis, consider the two-dimensional MDS layout of the justices and the two circles of influence (recall \S\ref{sec:fifth}) emanating from the majority and minority focal points, as depicted in Figure \ref{fig:fifthcase}.  One of the most striking properties exhibited here is that the majority focal point and the minority focal point are extremely close to each other and to the center of the Court.  What seems to distinguish the majority justices from the minority justices is not a left-right divide nor even a North-South divide on the somewhat mysterious second dimension, but rather the centrality of their locations.  Indeed, for the most part the minority justices are the more ``extreme'' justices in a two-dimensional, radial sense---that is, they tend to be further from the center of the court, whereas the majority justices are primarily clustered closer toward the center.  

\begin{figure}
\begin{center}
\includegraphics[trim={0.0in 0.8in 0.0in 0.0in},clip,scale=0.4]{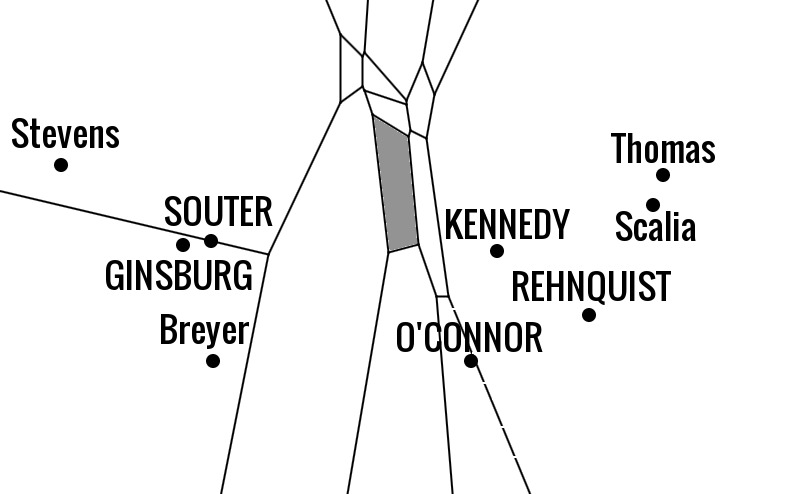}\\
\end{center}
\caption{The 5th order Voronoi diagram for the two-dimensional MDS layout of the justices on the 1994--2004 natural court.  The majority justices in \emph{Rogers v. Tennessee} are in upper-case, and the convex region they correspond to in the Voronoi diagram is filled in solid.  This solid gray trapezoidal region can be interpreted as an estimate of where this case lies in this two-dimensional MDS ideal space.}
\label{fig:Vorcase}
\end{figure}

This intriguing situation suggests that this voting coalition is not a half-plane coalition---it would be, were it not for Breyer's minority vote preventing an appropriately placed horizontal line from separating a Northern minority from a Southern majority---but that it is a Voronoi coalition whose corresponding convex region is located near the center of the court.  Both of these assertions are indeed verified by our algorithms; see Figure \ref{fig:Vorcase} for the 5th order Voronoi diagram showing a prominent North-central trapezoidal region uniting the majority justices.

Turning back to the circles of influence in Figure \ref{fig:fifthcase}, we see an intricate assessment of the decisive fifth vote in this case.  First, recall that the mean justice for this natural court is Kennedy (Table \ref{tab:rates}), but the term-specific mean justice for the years surrounding this case vacillates between Kennedy and O'Connor (Table \ref{tab:mJ}).  Indeed, Kennedy and O'Connor both orbit around the center of the Court without one or the other consistently taking the dominant role in these years.  The fifth vote as determined by proximity to the minority focal point (i.e., the majority justice most tempted by the minority outlook) is Kennedy, as is seen by his proximity to the square {\tiny $\blacksquare$} in Figure \ref{fig:fifthcase}, which accords with his status as a frequent swing vote.  

However, the fifth vote as determined by proximity to the majority focal point (i.e., the majority justice least convinced of the majority outlook) is Ginsburg, since she is the furthest majority justice from the triangle {\small $\blacktriangle$} in Figure \ref{fig:fifthcase}.  This makes sense from a political perspective, as Ginsburg is viewed as a traditionally liberal justice and so would be tempted by the prodefendant's rights viewpoint, as Edelman et al.\ point out.  But it also makes sense from a geometric perspective: Ginsburg's placement fairly far to the left of the Court means she is rather far from the centrally located ideological focal point of this case.

In summary, we see that this particularly perplexing oddball coalition is nicely conceptualized by the Voronoi spatial model.  The majority consists of a mixture of both liberal and conservative justices, and while it mostly comprises the more centrist justices among these, a one-dimensional liberal-to-conservative perspective fails here since the minority vote of Breyer is more central than the majority vote of Ginsburg.  What distinguishes Breyer from the majority is his vertical distance from the center of the court.  One could interpret the vertical axis as in \cite{2ndDimension} to try to explain Breyer's vote, but even without a legalistic interpretation we still see a compelling story geometrically: the five majority justices have a ``common ground,'' so to speak, which is the trapezoidal Voronoi region depicted in Figure \ref{fig:Vorcase} just north of the center of the court.  Thus, the majority is not the five most central justices in any one-dimensional scale, nor even the five most central justices in our two-dimensional MDS space (since that would include Breyer and lose Ginsburg), but it is the five justices most closely clustered around an idealogical point just north of the center of the Court.


\section{Conclusion}

The main motivation for this study is the observation of Edelman that Supreme Court majorities which deviate from the expected liberal-conservative schism lack a robust theoretical framework \cite[p. 569]{DimSupCourtEdelman}.  Progress has been made by various authors since Edelman made this remark, and a significant step forward was achieved in particular by Fischman and Jacobi when they uncovered some appealing structure in the Court's second dimension that helps explain certain oddball coalitions \cite{2ndDimension}.  This important work of Fischman and Jacobi further motivates the project undertaken in the present paper, which is a deeper geometrical exploration of the two-dimensional voting behavior of the Supreme Court.  We employ notions from mathematics such as $k$-sets, higher order Voronoi diagrams, and arrangements of circles to investigate the rich interactions between Fischman and Jacobi's two dimensions of the Court..  

This provides new spatial models for understanding swing votes and median justices, as well as spatial models for conceptualizing the distribution of 5-to-4 majorities on the Court.  The latter accord with historical voting records in approximately one third to half of the scenarios, thus making tremendous strides towards a comprehensive theory of oddball and ``unexpected'' Supreme Court rulings, while still leaving room for significant further improvements.  Our spatial methods, moreover, provide a collection of practical, computational tools for analyzing individual cases that can lead to new insight into the forces pulling the justices' votes in different directions and which can help uncover previously unnoticed idealogical common ground uniting various justices.


\end{document}